# Inventions on dialog boxes used in GUI
## A TRIZ based analysis

**Umakant Mishra**

Bangalore, India

http://umakantm.blogspot.in

**Contents**

# 1. Introduction

Dialog boxes are used in graphical user interfaces to display a few possible options based on a typical scenario. The dialog boxes are useful in case of displaying warnings, errors, confirmations etc. in special situations. A typical dialog box is displayed in a small window with some text message along with a few options for the user to select. A dialog box might display the attributes of an object such as the color, size and position etc. and allows the user to alter them.

However, there are certain difficulties associated in programming and implementing a conventional dialog box. For example:

- Dialog boxes take intensive care for programming as each dialog box is intended to handle a critical situation. Including hundreds of dialog boxes requires a lot of programming effort.

- The contents of the dialog box are generally hard coded. If there is any change in the program flow, the concerned dialog boxes may need to be revised accordingly.

- The display of the dialog box occupies substantial screen space thereby blocking the visibility of valuable information on the screen.

- The traditional dialog box mechanism cannot manage displaying multiple dialog boxes at a time.

- The modal dialog boxes do not allow the user to work on the application until the user explicitly closes them. However, the modeless dialog boxes are difficult to control as they can open in multiple and clutter the screen.

Thus, an ideal dialog box should be deprived of the above mentioned and other drawbacks. The dialog box should not obscure the screen. The user should be able open multiple dialog boxes but without obscuring the screen. The dialogs may have user configurability, previewability and other advanced features.

# 2. Inventions on dialog boxes

This article analyses 5 interesting inventions on dialog boxes selected from US Patent database. Each invention tries to overcome some limitations of a conventional dialog box and provides some positive features. Each solution is also analyzed from a TRIZ perspective.

## 2.1 Dynamic dialog box facility (5821932)

### Background problem
The dialog boxes typically display some message on the screen. The messages are normally informative and sometimes seek an input from the user. The programmer or developer spends a lot of time and effort to build a dialog box. As a dialog box is quite frequently used in an application, the development needs a lot of programming effort.

### Solution provided by the invention
William Pittore invented a dynamic dialog box (Patent 5821932, Assigned to Sun Microsystems, Oct 98), which provides easy modification of the information to be provided through the dialog box.

The new system includes a information type source file, information value source file and a dynamic dialog box processor. The dynamic dialog box processor uses the text entries from the information type source file to display the dialog box and receives a value depending on the information value file. Using this method, the operator can easily modify the dialog boxes by editing the information files, without any programming effort.

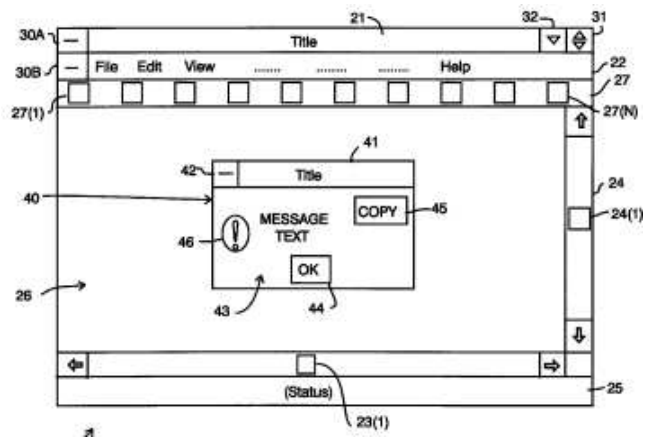

### TRIZ based analysis
The invention isolates the messages of dialog boxes from the dialog box engine and stores in a separate message file **(Principle-2: Taking out).**

The dialog messages in the message file can be easily edited by the operator **(Principle-15: Dynamize).** The dynamic dialog box processor uses the message text and generates dialog boxes for display to an operator **(Principle-5: Merging).**

## 2.2 Method of providing preview capability to a dialog box (6061059)

**Background problem**

The dialog boxes are used in user interfaces to display specific messages to the user and get the user feedback to the program control. The program does the relevant operation depending on the feedback from the dialog box and the operation is done permanently. It will be nice to have previewable dialogs, which allows the user to see the effects of the dialog control before choosing the dialog.

**Solution provided by the invention**

Taylor et al. disclosed a method (Patent 6061059, assigned to Adobe Systems Incorporated, may 2000) of providing a preview capability to dialogs. The previewable dialog will process the commands produced by the dialog's control to provide a preview capability.

According to the invention, the preview mechanism includes a do and undo mechanism kept in dialog commands buffer. The method bundles the commands in the dialog command queue into a single macro command, so that the whole set of commands can be done and undone.

The previewable dialog has the advantage that the developer can just add and remove the preview capability to a dialog by simply adding and removing the preview control. The developer need not be aware of how the preview control is implemented.

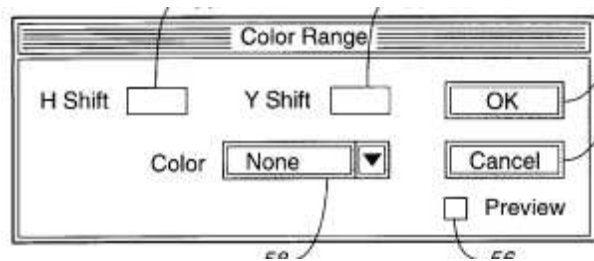

**TRIZ based analysis**

The invention provides a preview before the user finally decides to select an option **(Principle-27: Cheap and disposable).**

This functionality prevents user from taking poor/ improper decisions **(Principle-8: Counterweight).**

The preview mechanism includes a series of do and undo functions related to the specific dialog box operations. The functions are done for preview and undone again to bring the status to normal. **(Principle-13: Reversing).**

## 2.3 System and method for displaying multiple dialog boxes in a window display (6091415)

**Background problem**

A dialog box is used to conveniently display predefined messages and dialogs in a graphical user interface. However, a dialog box typically displays only one dialog in a window. But some special cases (e.g., electronic organizers, dictionaries etc.) may need to display multiple dialog boxes to display all pertinent information.

**Solution provided by the invention**

Patent 6091415 (invented by Chang et al., assigned by Inventec Corporation, issued July 2000) discloses a method of displaying multiple dialog boxes in a window. The invention displays various types of information in a family of dialog boxes. These dialog boxes in the same family are cascaded together to occupy less space in the display screen.

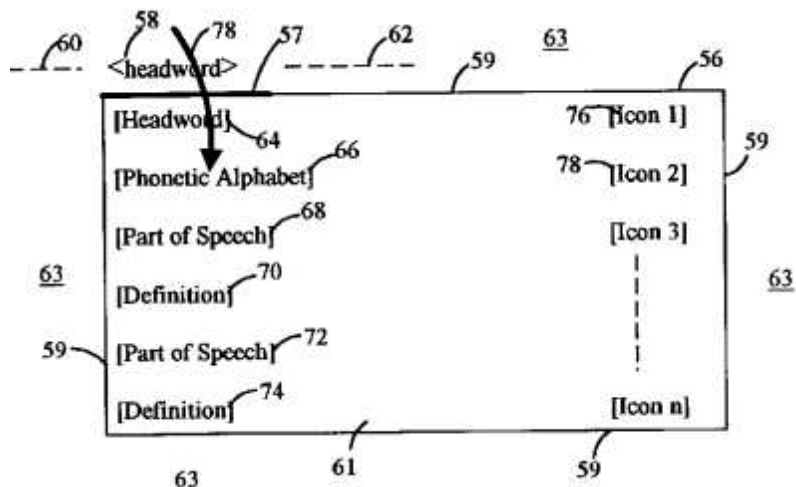

This method can be effectively used to display a plurality of translated words, synonyms, antonyms, audio pronunciations, bilingual translation, exemplary sentences etc. in a plurality of dialog boxes. As in case of a bilingual dictionary a user may need to look up many words before he or she understands the meaning of the sentence, it is helpful if the meaning of each word remains displayed simultaneously on the screen.

**TRIZ based analysis**

The invention displays many events simultaneously in multiple dialog boxes **(Principle-5: Merging).**

The invention cascades the dialog boxes of the same family together so that they occupy less screen space **(Principle-7: Nested doll).**

The invention provides a method of displaying menus or dialog boxes automatically without clicking the mouse or striking a key **(Principle-25: Self service).**

## 2.4 Apparatus and method for controlling dialog box display and system interactivity in a computer-based system (6118451)

### Background problem

The conventional dialog boxes provide two modalities, viz., modal and modeless. The modal dialog boxes do not allow users to operate on the application unless explicitly closed by the user. On the other hand modeless dialog boxes allow user to access and operate other elements of the application interface. But both of them have their disadvantages.

A modal dialog box is problematic when the user needs to access other system components prior to entering the control information into the active dialog box. When the user is forced to close the dialog box without knowing what he should actually enter, the system may produce an undesirable result.

The modeless dialog box overcomes the above drawbacks of the modal dialog box, as the modeless dialog box does not restrict the user's access to other GUI elements. But the modeless dialog boxes can open multiple dialog boxes and clutter the display screen. Therefore, the dialog box should have a desired degree of interactivity without cluttering the display.

### Solution provided by the invention

Patent 6118451 (invented by Alexander, assignee Agilent Technologies, assigned Sep 2000) provides a method that overcomes the above drawbacks of conventional dialog box management techniques. According to the invention the dialog box modalities can be modal or semi-modeless. The modal dialog box provides a first degree of display clarity. With the semi-modeless dialog provides a second degree of display clarity approximately the same as the first degree of display clarity and a second extent of system interactivity beyond the active dialog box that is greater than the first extent of system interactivity.

According to the invention a display controller controls the display of and interactivity with the associated dialog boxes. The dialog box control system determines the modality of the dialog box and closes the open dialog boxes not having a predetermined relationship with the selected dialog box. This enables the invention to provide an associated degree of display clarity and an extent of system interactivity associated with the dialog box modality.

### TRIZ based analysis

The dialog box should be modal to not allow too many dialogs to clutter the screen, at the same time, it should be modeless to allow user to access other elements of GUI (**Contradiction**).

The invention provides a new modality "semi-modeless" which has advantages of both the modal and modeless dialog boxes **(Principle-16: Partial or excessive action)**.

## 2.5 Message box facility for graphical user interface for computer system video display (6414699)

### Background problem
The graphical user interface often displays error messages and dialog boxes in response to various user operations. The user also interacts through those message boxes and dialog boxes to provide a suitable response to the computer. However, one problem with such message boxes is that the content of the message boxes are vanished as soon as the user responds to it. Unless the user writes down the message on a paper he cannot later know which conditions caused the message to appear.

### Solution provided by the invention
Patent 6414699 (invented by Pittore, assigned by Sun Microsystems, issued Jul 2002) provides a method of retaining the message information after the message boxes have been removed.

The invention includes a actuable copy enable facility and an actuable message box removal facility. When the message box is displayed it actuates a copy enable facility which copies the message text from the message box to the common buffer (such as windows clipboard) and an actuable message box removal facility that closes the message box dialog. This method stores the message in the buffer/ clipboard to be viewed later by a clipboard viewer or text editor.

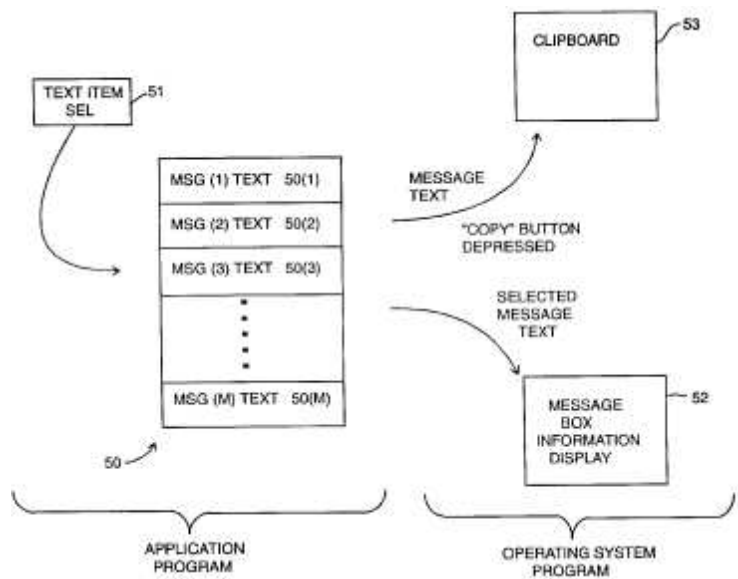

### TRIZ based analysis
The invention copies the message displays in the message box to the clipboard for later viewing **(Principle-26: Copying).**

## 3. Summary

In the above illustrations we found some interesting inventions on improving a conventional dialog box. One invention provides a preview capability to the dialog box, so that the user can preview the consequences of selecting an option without actually selecting it. Another invention provides a mechanism to manage multiple dialog boxes. Another invention provides a semi-modeless dialog box, which consists of the features of both modal and modeless dialog-boxes. There is enough scope for inventing many more features of a dialog box to make it much more powerful.